\documentclass[12pt,preprint,useAMS,usenatbib]{emulateapj}
\usepackage{graphicx}
\usepackage{setspace}
\usepackage{natbib}
\usepackage{color}
\usepackage{amsmath,amssymb}

\begin{document}

\title{Stellar mass-gap as a probe of halo assembly history and concentration: youth hidden among old fossils}
\author{A. J. Deason\altaffilmark{1,4}, C. Conroy\altaffilmark{1,5}, A. R. Wetzel\altaffilmark{2}, J. L. Tinker\altaffilmark{3}}

\altaffiltext{1}{Department of Astronomy and Astrophysics, University
  of California Santa Cruz, Santa Cruz, CA 95064, USA; alis@ucolick.org}
\altaffiltext{2}{Department of Astronomy, Yale University, New Haven, CT 06520, USA}
\altaffiltext{3}{Center for Cosmology and Particle Physics, Department of Physics, New York University, New York, NY 10013, USA}
\altaffiltext{4}{Hubble Fellow}
\altaffiltext{5}{Alfred P. Sloan Fellow}
\date{\today}

\begin{abstract}
We investigate the use of the halo mass-gap statistic --- defined as the logarithmic difference in mass between the host halo and its most massive satellite subhalo --- as a probe of halo age and concentration. A cosmological $N$-body simulation is used to study $N \sim 25, 000$ group/cluster sized halos in the mass range $10^{12.5} < M_{\rm halo}/M_\odot < 10^{14.5}$. In agreement with previous work, we find that halo mass-gap is related to halo formation time and concentration. On average, older and more highly concentrated halos have larger halo mass-gaps, and this trend is stronger than the mass-concentration relation over a similar dynamic range. However, there is a large amount of scatter owing to the transitory nature of the satellite subhalo population, which limits the use of the halo mass-gap statistic on an object-by-object basis. For example, we find that 20\% of very large halo mass-gap systems (akin to ``fossil groups'') are young, and have likely experienced a recent merger between a massive satellite subhalo and the central subhalo. We relate halo mass-gap to the observable stellar mass-gap via abundance matching. Using a galaxy group catalog constructed from the Sloan Digital Sky Survey Data Release 7, we find that the star formation and structural properties of galaxies at fixed mass show no trend with stellar mass-gap. This is despite a variation in halo-age of $\approx 2.5$ Gyr over $\approx 1.2$ dex in stellar mass-gap. Thus, we find no evidence to suggest that the halo formation history significantly affects galaxy properties. 

\end{abstract}

\section{Introduction}

Old dark matter halos don't necessarily host old galaxies. An old halo accumulates most of its mass at early times in the Universe, and experiences little recent mass growth (see e.g. \citealt{wechsler02}). Whereas, an old galaxy has stopped (or has very little) stellar mass growth via star formation. The relation between halo age and galaxy age remains uncertain, and much work has been devoted to understand their connection (e.g. \citealt{kauffmann04}; \citealt{abbas06}; \citealt{yang06}; \citealt{zhu06}; \citealt{blanton07b}; \citealt{croton07}; \citealt{li08}; \citealp{tinker08a, tinker11}; \citealt{wang08}; \citealt{skibba09}; \citealt{cooper10}; \citealt{hearin13b}). Despite this large body of work, there is little, if any, compelling evidence that galaxy age correlates with halo age. 

An obvious difficulty in relating galaxy age to halo age is how to determine the age of a halo. Several age indicators exist for galaxies that relate to their star formation activity (e.g. the strength of the 4000 \AA\ break, or the equivalent width of Hydrogen Balmer lines). While such age-indicators suffer from their own problems (e.g. age-metallicity degeneracy), they can at least give a qualitative indication of whether or not a galaxy is actively forming stars. However,  there is no direct observational tool that can tell us the age of a dark matter halo. In simulations, we are privy to the full mass-accretion history of a halo, whereas observationally we are limited to a single redshift-snapshot of the luminous material belonging to each halo.

The concentration of a dark matter halo is strongly linked to when it formed; older halos form at early times when the Universe was denser, and hence they are generally more concentrated than later forming halos\footnote{Note, however, that concentrations can be uncertain or biased for dynamically unrelaxed halos (e.g. \citealt{maccio08}), and it has been suggested that concentration is only a good age indicator for relaxed systems (see e.g \citealt{wong12})}. Indeed, the well-known mass-concentration relation at a given epoch is a consequence of less massive halos forming, on average, at earlier times than more massive halos (see e.g, \citealt{nfw}; \citealt{bullock01}; \citealt{wechsler02}). The relation between halo mass and concentration has received a great deal of attention, both theoretically and observationally. However, despite being a fundamental prediction of $\Lambda$-Cold Dark Matter ($\Lambda$CDM), it remains to be seen whether observations agree with the theoretical predictions. For example, high mass cluster halos appear to be over-concentrated relative to $\Lambda$CDM predictions (e.g. \citealt{schmidt07}; \citealt{comerford07}; \citealt{hennawi07}; \citealt{broadhurst08}; \citealt{oguri12}), while there is some evidence that isolated, late-type galaxy halos are under-concentrated (e.g. \citealt{kassin06}; \citealt{dutton07}; \citealt{gnedin07}). However, observational constraints on halo mass and concentration are fraught with systematic influences (e.g. selection effects), and there are still relatively few measures of halo concentration, especially on the group-mass scale. 

Perhaps the most direct observational link to dark matter halos, particularly on the group/cluster mass scale, is via their satellite galaxy populations. For example, the radial distribution of the satellites approximately follows the dark matter profile, especially at large radii (e.g. \citealt{sales07}; \citealt{klypin11}; \citealt{budzynski12}; \citealt{tal12}), and satellite galaxies have often been used as dynamical tracers to constrain the total mass of halos (see e.g. \citealt{zaritsky97}; \citealt{prada03}; \citealt{conroy05}; \citealt{watkins10}; \citealt{deason13}). Furthermore, It has been recognized by several authors that the difference in mass (or luminosity) between the most-massive satellite and central galaxy may have an important link to the mass-assembly (and hence age) of a galaxy halo (e.g., \citealt{donghia05}; \citealt{milo06}; \citealt{beckmann08}; \citealt{dariush10}; \citealt{hearin13a}; \citealt{wu13}).

The potential link between halo age and ``magnitude gap", defined as the difference in absolute magnitude between the two most massive members of a galaxy group or cluster, has been appreciated by several authors in the context of fossil groups (e.g. \citealt{ponman94}; \citealt{jones03}; \citealt{donghia05}; \citealt{milo06}; \citealt{vandenbosch07}; \citealt{beckmann08}; \citealt{yang08}; \citealt{dariush10}). Fossil groups are generally defined as X-ray luminous galaxy groups where the majority of the mass is contained within the central galaxy, and there is a dearth of massive companion satellites. It has been suggested that these ``fossils" represent old halos which built up the majority of their mass at early times. Indeed, comparisons with numerical simulations have shown that large magnitude gap galaxy groups are more common amongst older halo populations. However, \cite{beckmann08}, \cite{dariush10} and \cite{cui11} caution that the single redshift snapshot of magnitude gap can lead to misleading definitions of fossils due to the transitory nature of the satellite population, which leads to a large variation of magnitude gap with time. Furthermore, while the central galaxies of fossil groups are often found to be ``red and dead", some observational studies find evidence for recent activity in these systems (e.g. \citealt{hess12}). 

In this study, we investigate the use of halo mass-gap --- defined as the logarithmic difference in mass between the host halo and its most massive subhalo --- as a probe of halo age and concentration for group/cluster scale halos. We employ a large statistical sample of simulated halos, which allows us to address both the median trends and associated scatter. We find that the scatter caused by the transitory nature of the halo mass-gap statistic limits its use on an individual halo-by-halo basis. However, despite the scatter, we demonstrate that the link between halo mass-gap and concentration is stronger than the mass-concentration relation, and may be a more useful, and observationally accessible, test of $\Lambda$CDM. Using a group catalog constructed from the Sloan Digital Sky Survey (SDSS) Data Release 7 (DR7), we find that there is no relation between stellar mass-gap (and therefore halo age) and the star formation and structural properties of galaxies at fixed mass. This result argues against a strong relation between galaxy age and halo age.

This paper is arranged as follows. In Section 2 we briefly describe the simulations employed in this work, and we relate halo mass-gap to halo age and concentration. In Section 3 we use a group catalog to relate observed galaxy properties to stellar mass-gap. We discuss the implications of our results in Section 4, and finally our main findings are summarized in Section 5. For all calculations we assume a flat, $\Lambda$CDM cosmology of ($\Omega_m$, $\sigma_8$, $\Omega_b$, $n_s$, $h_0$) = (0.27, 0.8, 0.046, 0.95, 0.7), and a \cite{chabrier03} stellar initial mass function. 

\section{Insights from simulations}
\subsection{Simulation details}
\label{sec:sims}

\begin{figure*}
    \centering
    \includegraphics[width=14cm, height=11.2cm]{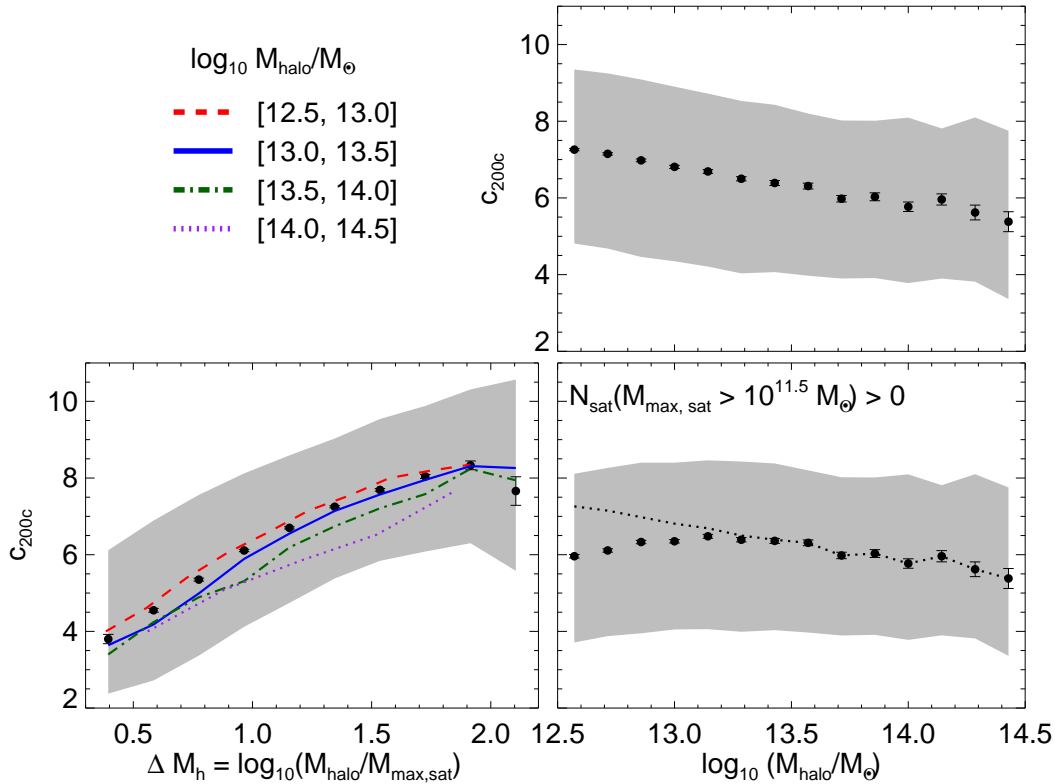}
    \caption[]{\small \textit{Top-right panel:} The mass-concentration relation for halos with masses $10^{12.5} < M_{\rm halo}/M_\odot < 10^{14.5}$. The black points with error bars show the median values, and the gray shaded region indicates the 68\% scatter. \textit{Bottom-right panel:} The mass-concentration relation for groups with at least 1 satellite subhalo, with a minimum satellite subhalo mass of $M_{\rm sat} = 10^{11.5} M_\odot$ ($M^*_{\rm sat} \sim 10^{9.7}M_\odot$). This restriction biases the lower mass halos against higher concentrations, and the resulting mass-concentration relation is essentially flat. The dotted line shows the mass-concentration relation for the full catalog (including isolated halos) for comparison. \textit{Bottom-left panel:} The relation between halo concentration and halo mass-gap, where halo mass-gap is defined as the logarithmic difference in mass between the parent halo and its most massive satellite subhalo within $r_{\rm halo}$, $\Delta M_h=\mathrm{log}_{10}(M_{\rm halo}/M_{\rm max, sat})$. The correlation is considerably stronger than the halo mass-concentration relation in this halo mass range. The dashed red, solid blue, dot-dashed green and dotted purple lines show the median values for different halo mass bins. All mass halos show the same correlation between concentration and halo mass-gap, although lower mass halos are biased towards higher concentrations.}
   \label{fig:conc_dmass}
\end{figure*}

The simulation that we employ is described in detail by \cite{wetzel13} and \cite{white10}. In brief, we use a dissipationless $N$-body simulation with flat $\Lambda$CDM cosmology ($\Omega_m=0.274$, $\Omega_b=0.0457$, $h = 0.7$, $n = 0.95$ and $\sigma_8 = 0.8$), which has a particle mass of $1.98 \times 10^8 M_\odot$ and Plummer equivalent smoothing of $2.5$ $h^{-1}$ kpc in a 250 $h^{-1}$ Mpc box. This allows us to achieve both significant volume as well as sufficiently high resolution to robustly track subhalos with masses $\gtrsim 10 ^ {11} M_\odot$ at infall.

The identification and tracking of halos and subhalos is described in \cite{wetzel13}. In brief, we identify halos using the Friends-of-Friends (FoF) algorithm (\citealt{davis85}) with a linking length of $b = 0.168$ times the average particle spacing. Within halos, we identify subhalos as overdensities in phase space through a 6-dimensional FoF algorithm (FoF6D). We keep all objects with at least 50 particles, and we define the center and velocity via the most bound particle. We define a ``central'' subhalo as being the most massive subhalo (at the minimum of the potential well) in a newly formed halo, and a subhalo retains its central demarcation until it falls into another halo, becoming a ``satellite''. We assign to each subhalo a ``peak'' mass, $M_{\rm peak}$, as given by the maximum halo mass that it ever had as a central subhalo, motivated by the expected correlation of this quantity with galaxy stellar mass (see Section \ref{sec:stellar}). Following \cite{wetzel10}, we define a satellite subhalo as destroyed (either by merging with the central subhalo or otherwise disrupted) if it falls below $0.007 M_{\rm peak}$. For both halos and subhalos, our catalog is based on a 50 particle limit. However, we impose a $M_{\rm peak} > 10^{11}M_\odot$ limit on both halos and subhalos, driven by the ability to resolve satellite subhalos properly. For halos, we only use objects with $> 500$ particles. For subhalos, we only use objects that had at least 500 particles at infall, resolving them down to 50 particles in instantaneous/bound mass.

To be consistent with our observational galaxy group catalog (see Section \ref{sec:groups}), we define the halo radius ($r_{\rm halo}$) as the radius within which the mean density is 200 times the background matter density. We measure halo concentration by performing (unweighted) Navarro-Frenk-White (NFW; \citealt{nfw}) profile fits to the radial dark matter density distribution. We define concentration according to $c_{200c}=r_{200c}/r_s$, where $r_s$ is the scale radius and $r_{200c}$ ($< r_{\rm halo}$) is the radius within which the mean density is 200 times the critical density.

In this work, we consider group/cluster scale halos at $z=0.05$ with masses in the range $10^{12.5} < M_{\rm halo}/M_\odot < 10^{14.5}$. In this mass-range we have sufficient resolution and statistics to explore the satellite subhalo population of each halo. There are $N=25, 076$ halos in this mass range which have at least one satellite subhalo. 

In the following sections we explore the relation between halo age and the difference in mass between the parent halo and its most massive satellite subhalo.\footnote{Note that throughout this work we define ``halo mass-gap'' as the logarithmic difference in mass between the host halo and its most massive satellite subhalo, and we use ``stellar mass-gap'' to define the logarithmic difference in stellar mass between the central galaxy and its most massive satellite galaxy.}

\subsection{Halo mass-gap and concentration}

\begin{figure*}
\begin{minipage}{0.5\linewidth}
  \centering
  \includegraphics[width=8.5cm, height=6.8cm]{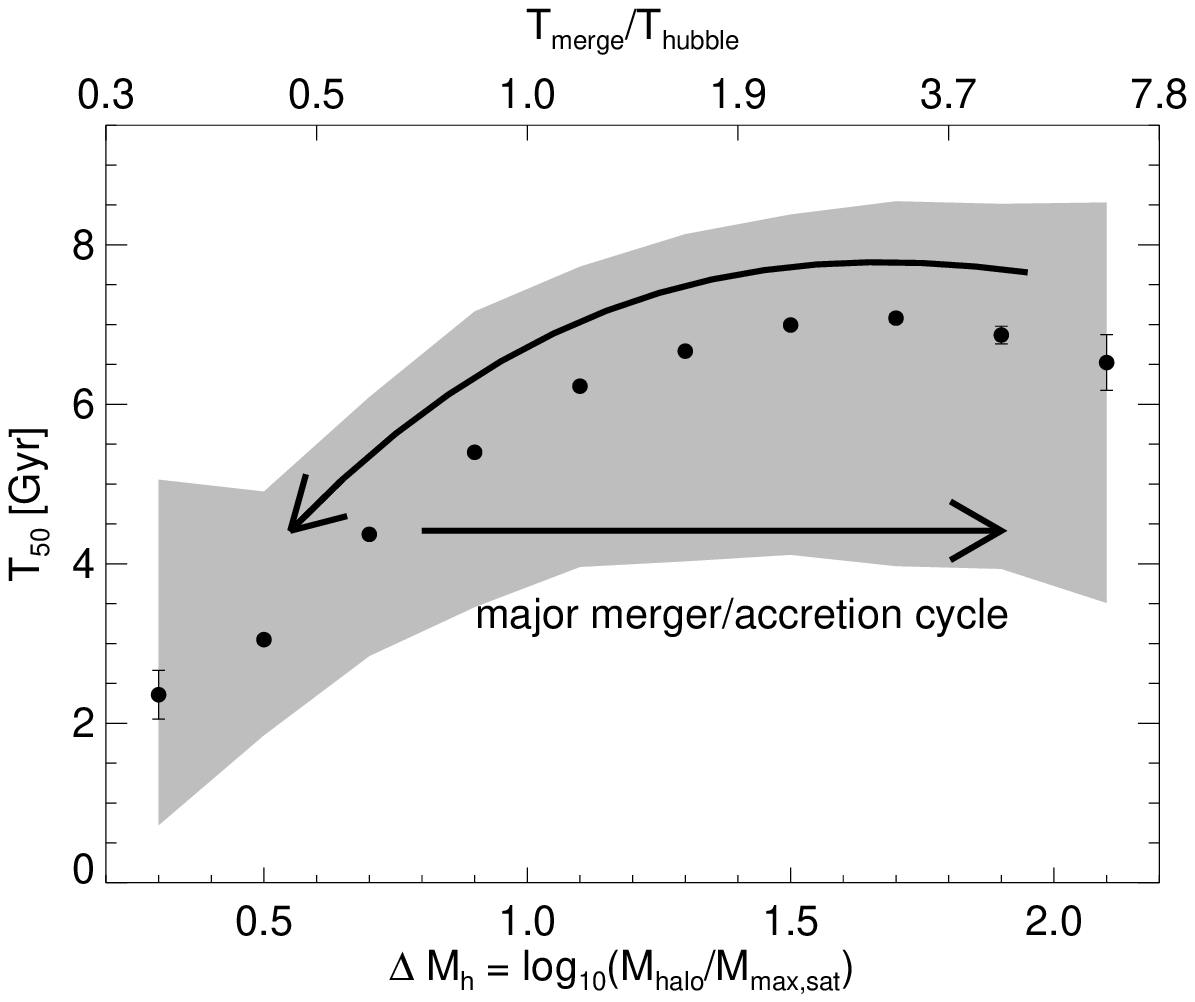}
 \end{minipage}
 \begin{minipage}{0.5\linewidth}
   \includegraphics[width=8.5cm, height=6.8cm]{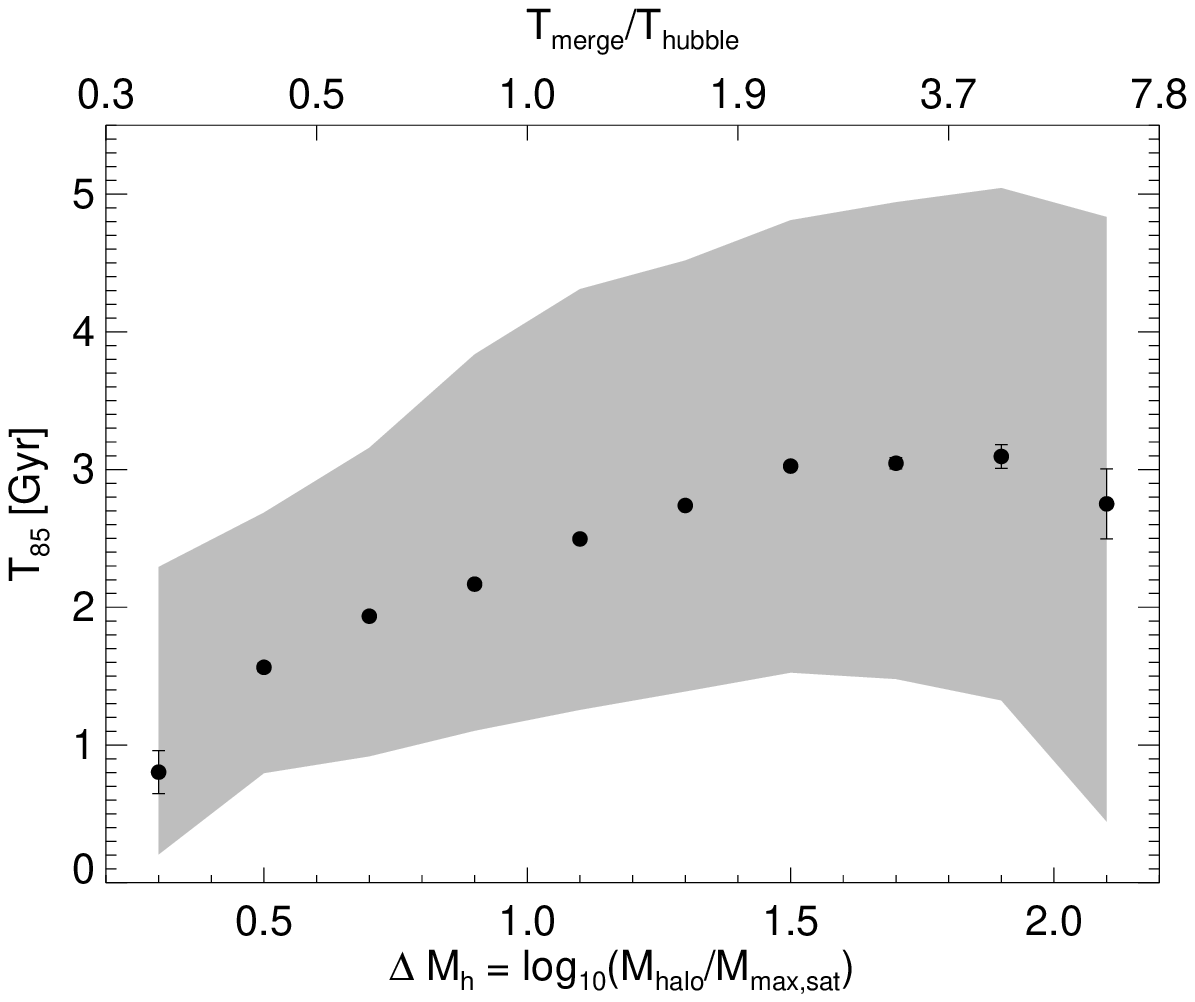}
 \end{minipage}
 \caption[]{\small Halo age -- defined as the lookback time when 50\% (left panel) or 85\% (right panel) of the $z=0$ halo mass of the main progenitor is in place -- against halo mass-gap. The black points with error bars show the median values, and the gray shaded region indicates the 68\% scatter. The large scatter at high halo mass-gaps is partly due to recent major mergers (example paths shown by arrows). This emphasizes the transitory nature of the halo age--halo mass-gap relation. The top axis indicates the approximate satellite subhalo lifetime based on $\Delta M_h$. This is computed from the calibration derived by \cite{wetzel10}. Systems where $T_{\rm merge}/T_{\rm hubble} \ll 1$ will be transient on average.}
 \label{fig:dmass_age}
\end{figure*}

A well-known relation exists between halo mass and concentration (see e.g. \citealt{gao08}; \citealt{maccio08}; \citealt{zhao09}; \citealt{klypin11}; \citealt{prada12}). Lower mass halos typically form at earlier times than higher mass halos. As the density of the Universe is higher at early times, this leads to lower mass halos having, on average, higher concentrations than more massive halos. The top right-panel of Fig. \ref{fig:conc_dmass} shows the mass-concentration relation for halos in the mass range $10^{12.5} < M_{\rm halo}/M_\odot < 10^{14.5}$. The black dots with error bars indicate the median values and the gray shaded region shows the 68 \% scatter. 

Despite the widespread attention to this relation, the variation in concentration over a wide range in halo mass is fairly weak, and there is considerable scatter. The bottom-left panel shows that there is a much stronger correlation between halo concentration and \textit{halo mass-gap}. Here, we define halo mass-gap as the logarithmic difference in mass between the parent halo and its most massive satellite subhalo within $r_{\rm halo}$: $\Delta M_h \equiv \mathrm{log}_{10}\left(M_{\rm halo}/M_{\rm max, sat}\right)$. Note that these panels cover roughly the same dynamic range in halo mass and halo mass-gap ($\sim 2$ dex). The dashed red, solid blue, dot-dashed green and dotted purple lines indicate the median values for group/cluster scale halos with $10^{12.5} < M_{\rm halo}/M_\odot < 10^{13}$, $10^{13} < M_{\rm halo}/M_\odot < 10^{13.5}$,  $10^{13.5} < M_{\rm halo}/M_\odot < 10^{14}$ and  $10^{14} < M_{\rm halo}/M_\odot < 10^{14.5}$ respectively. The same relation holds for each halo mass bin, but the lower mass halos are slightly more concentrated than higher mass halos. 

We note that for low halo mass-gap systems, the presence of large substructures could significantly affect the NFW profile fit. However, given that the correlation between concentration and halo mass-gap continues to much higher halo mass-gaps systems (i.e. $\Delta M_h > 1$), where the effect of substructure should be minimal, we can be confident that the apparent trend is not a consequence of a varying amount of substructure between low and high halo mass-gap systems. Furthermore, we show in Section \ref{sec:age} that a similar relation exists between halo-age and halo mass-gap, which is independent of our parametrization of the dark matter density profile.

The association between halo concentration and halo mass-gap can influence the mass-concentration relation. In the bottom-right panel of Fig. \ref{fig:conc_dmass} we show the mass-concentration relation, but only for halos with at least one satellite subhalo, with a lower mass threshold of $M_{\rm max, sat} > 10^{11.5}M_\odot$. This is the typical bias introduced when studying group catalogs with at least one satellite galaxy (with a typical stellar mass-threshold of $M^* \approx 10^{9.7}M_\odot$\footnote{Note that this threshold is not arbitrary but corresponds to the stellar mass limit of our group catalog (see Section \ref{sec:groups})}). This restriction has little affect on the high mass end, but significantly reduces the average concentration of the lower mass halos. leading to an essentially flat mass-concentration relation. This bias at the low mass end is due to the absence of halos with large halo mass-gaps, and hence higher concentrations.

For many years, observational studies have attempted to match the mass-concentration relation predicted by simulations. This has met with varying degrees of success, and at present there is no strong evidence that this relation holds in the real Universe (see e.g. \citealt{duffy08}; \citealt{oguri12}). However, perhaps this in unsurprising given the fairly weak trend ($c \propto M^{-0.1}$) and large scatter. Fig. \ref{fig:conc_dmass} suggests that a stronger trend should be apparent between concentration and halo mass-gap, and this may be easier to verify observationally. In particular, a wide range in halo mass is required to study the full dynamic range of the mass-concentration relation. This often means that inhomogeneous observations, using a wide-range of methods to measure concentration and with different selection biases, are used to probe the mass-concentration relation. However, only a small range in halo mass is required to probe the correlation between halo mass-gap and concentration. In Section \ref{sec:stellar}, we relate halo mass-gap to more accessible observational quantities.

\subsection{Halo mass-gap and age}
\label{sec:age}
Halo concentration is strongly related to assembly history; halos that accumulated most of their mass at early times are more concentrated than late forming halos. Thus, the correlation between halo mass-gap and halo concentration found in the previous section suggests that halo mass-gap may be a good proxy for halo ``age''. Unfortunately, the definition of halo age is ambiguous, and several different definitions are adopted in the literature\footnote{Note that halo concentration itself is often considered as an early time age indicator.}. In this work, we use two age indicators, defined as the lookback time when 50\% or 85\% of the present day halo mass was assembled in the main progenitor: $T_{50}=T(M=0.50M_0)$, $T_{85}=T(M=0.85M_0)$. $T_{50}$ is a ``canonical'' age indicator which is used widely in the literature, while $T_{85}$ represents mass growth within the last few Gyr (see \citealt{tinker11}) and relates more directly to timescales of stellar age that we can measure.

The relation between halo age and halo mass-gap can be explained by simple dynamical considerations. A massive satellite subhalo will sink to the center of its parent halo over a dynamical timescale. The earlier a halo is formed, the more likely it is that any massive satellite subhalo has been cannibalized by the parent halo, leaving only lower mass survivors (i.e. those satellites that are still distinct subhalos) with larger halo mass-gaps. The extreme example of this are the so-called ``fossil groups'', which have a massive central galaxy but are devoid of any massive satellite galaxies. The bottom-left panel of Fig. \ref{fig:conc_dmass} shows that there is no clear distinction between a fossil and a non-fossil group, but rather there is a continuous trend between halo concentration (or halo age) and halo mass-gap. 

\subsubsection{The transitory nature of the halo mass-gap statistic}

\begin{figure}
  \centering
    \includegraphics[width=8.5cm, height=15.3cm]{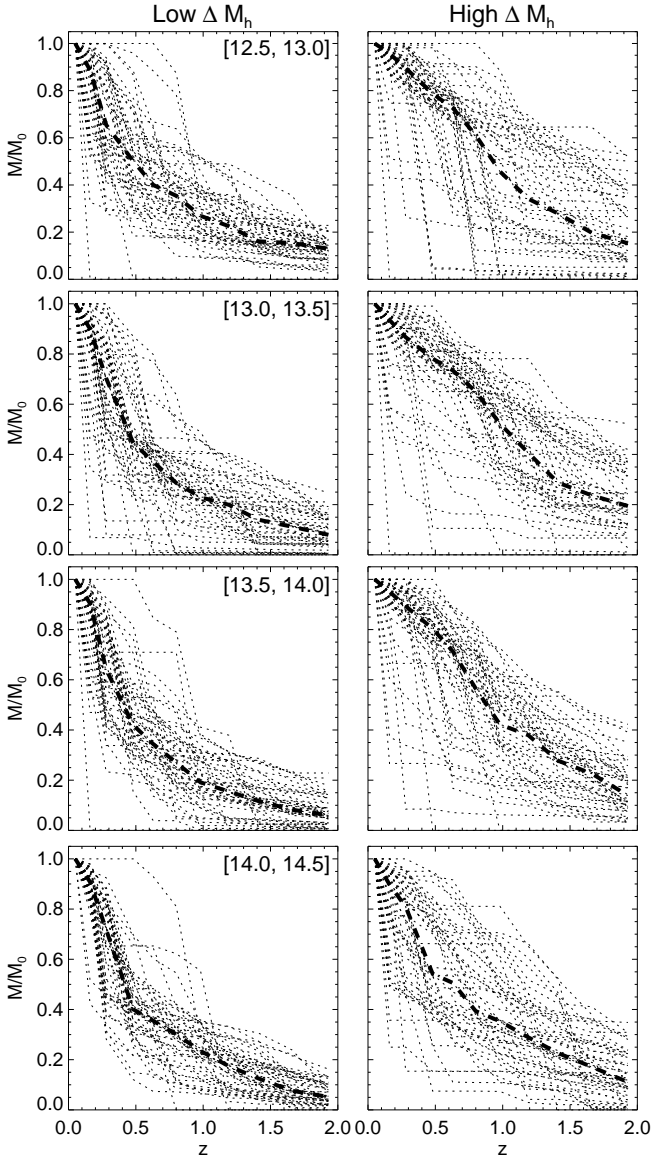}
    \caption[]{\small Mass accretion histories of 50 randomly selected halos with low (left panels) and high (right panels) halo mass-gaps. The thick dashed lines indicate the median mass accretion histories. Here, we define ``low'' ($\Delta M_h \lesssim 0.7$) and ``high'' ($\Delta M_h \gtrsim 1.6$) halo mass-gaps as the lowest/highest 10\% of the population. Different mass bins are shown in each row, and the numbers in brackets indicate the log halo mass ranges. A visual difference between low and high halo mass-gap systems is seen in the mass accretion histories: low halo mass-gap systems have undergone more recent mass growth, whereas high halo mass-gap systems experience more mass growth at early times.}
  \label{fig:mass_acc}
\end{figure}

In Fig. \ref{fig:dmass_age} we show the correlation between $T_{50}$, $T_{85}$ and halo mass-gap. As expected given the halo mass-gap--concentration relation, a corresponding association exists between halo-age and halo mass-gap. There is a strong correlation on average for both age indicators, but at large halo mass-gaps the relation flattens and the scatter increases. Note that this flattening, and increased scatter is also seen (to a lesser extent) in the relation between halo mass-gap and concentration.

We briefly mentioned earlier that the definition of halo age is somewhat ambiguous, so it is informative to directly show the mass-accretion histories of halos with different halo mass-gaps. In Fig. \ref{fig:mass_acc} we show the mass accretion histories of 50 randomly selected halos with low (left panels) and high (right panels) halo mass-gaps, and different mass bins are shown in each row. Here, we define ``low'' and ``high'' halo mass-gaps as the lowest/highest 10\% of the population. A clear difference in assembly history is evident between low and high halo mass-gap systems; very recent mass-growth is dominant for low halo mass-gap systems, and earlier mass-growth is more common for high halo mass-gap systems.

In a recent paper, \cite{diemer13} (see also \citealt{diemand07}; \citealt{cuesta08}) show that dark matter halos undergo ``pseudo-evolution'' of halo mass simply due to the redshift evolution of the chosen reference density (e.g. 200$\delta_m$). Thus, the mass accretion histories we show in Fig. \ref{fig:mass_acc} will have some contribution from pseudo evolution, which may affect the absolute values of halo age that we use in this work. However, as we are interested in the \textit{relative} differences in mass-accretion histories of halos at fixed halo mass, this should have little affect on our main results. Furthermore, as pointed out by \cite{diemer13}, pseudo mass evolution is more pronounced for lower mass halos ($M_{\rm halo} \lesssim 10^{12} M_\odot$) than the group/cluster scale halos considered here.

\begin{figure}
  \centering
\includegraphics[width=8.5cm, height=6.8cm]{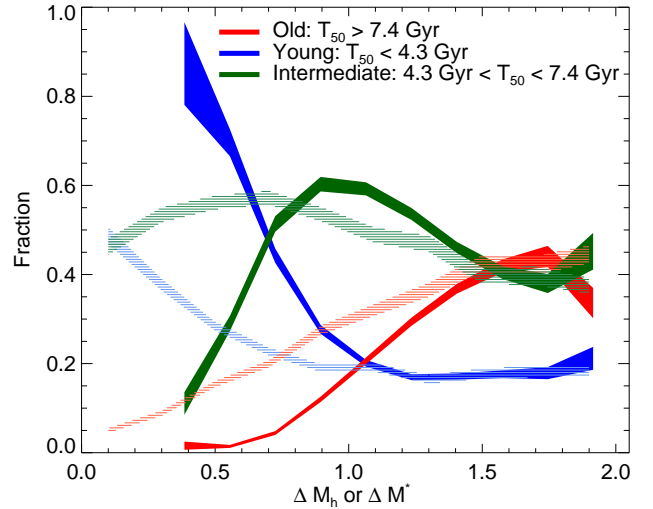}
  \caption[]{\small The fraction of young ($T_{50} < 4.3$ Gyr, red), intermediate ($4.3 <T_{50}/\mathrm{Gyr} < 7.4$, green) and old ($T_{50} > 7.4$ Gyr, blue) halos as a function of halo (and stellar) mass-gap. The solid filled regions are for halo mass-gap, while the line-filled regions are for stellar mass-gap (see Section \ref{sec:stellar} and Fig. \ref{fig:dmhalo_dmstar}). The fraction of old/young halos increases/decreases with halo mass-gap. However, there remains a significant fraction of young halos ($\sim 20\%$) with large halo mass-gaps.}
  \label{fig:fractions}
\end{figure}

\begin{figure*}
\begin{minipage}{0.5\linewidth}
  \centering
  \includegraphics[width=8.5cm, height=6.8cm]{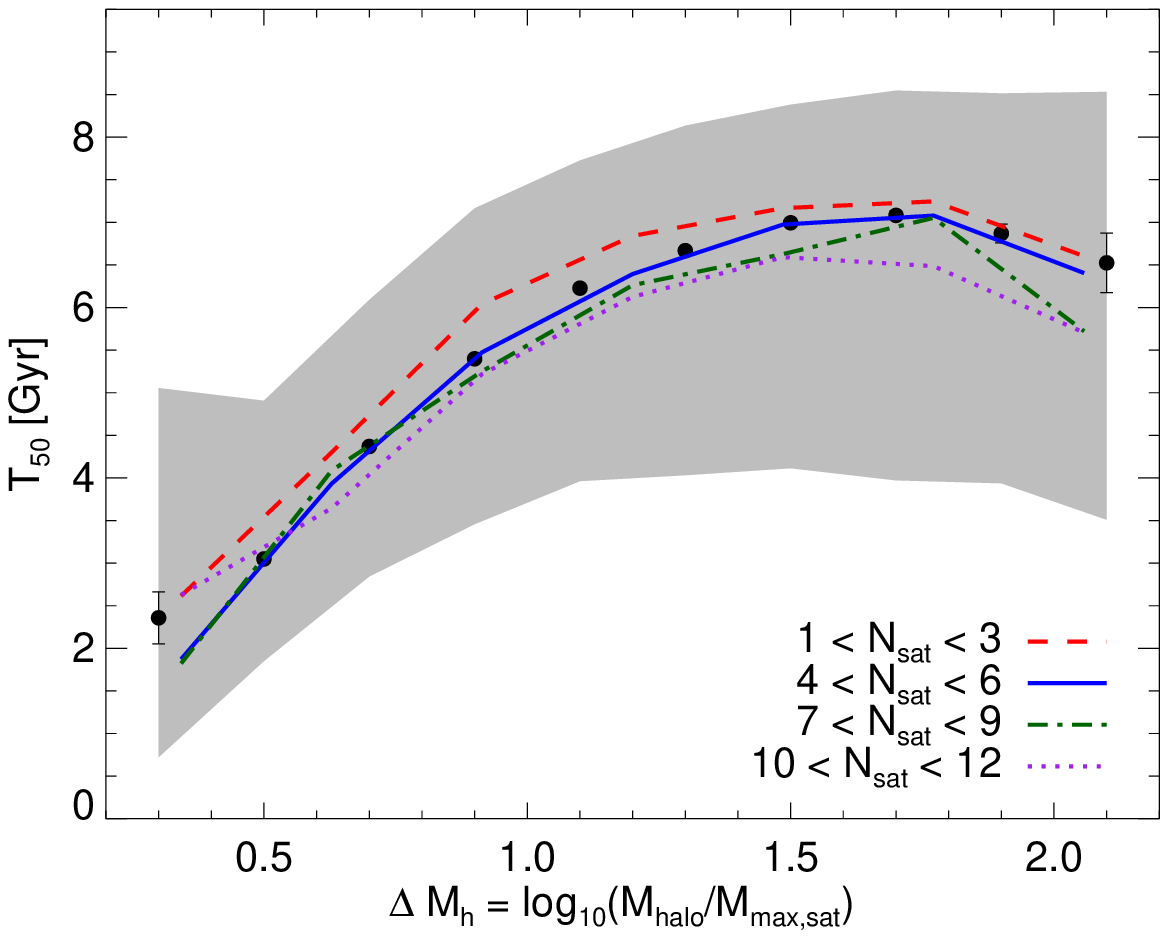}
 \end{minipage}
 \begin{minipage}{0.5\linewidth}
   \includegraphics[width=8.5cm, height=6.8cm]{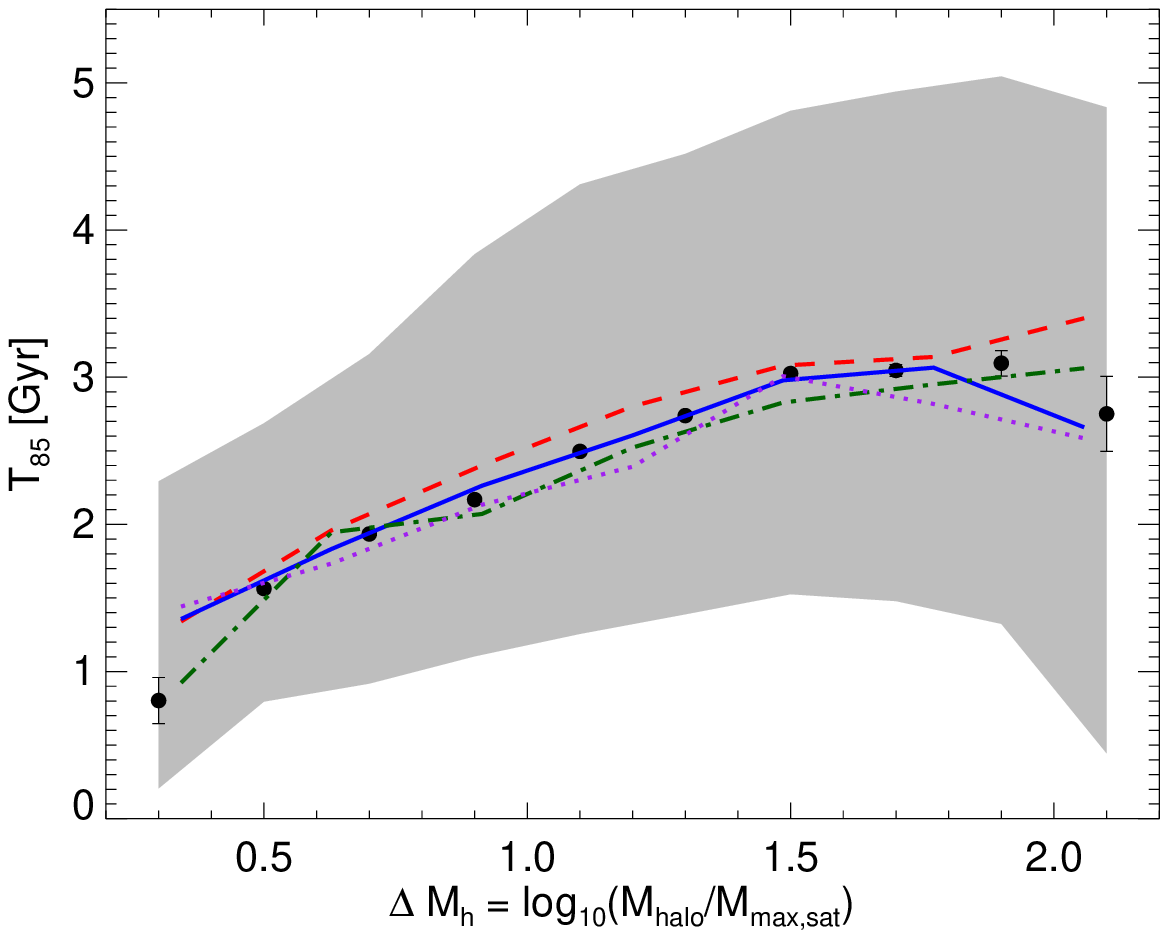}
 \end{minipage}
 \caption[]{\small \small Halo age ($T_{50}$ and $T_{85}$) against halo mass-gap, for fixed \textit{richness} bins. Halo age (as defined here) is a stronger function of halo mass-gap than group richness.}
 \label{fig:dmass_age_nfixed}
\end{figure*}

A visual inspection of halos with large halo mass-gaps ($\Delta M_h \sim 2$) shows that a significant number of these systems ($\sim 20\%$) have experienced a recent merger between the most massive satellite subhalo and the central subhalo. Thus, while at $z=0$ they may appear as ``isolated fossils'', they have not undergone a quiescent evolution, as is generally assumed for large halo mass-gap systems. More generally, the majority of the scatter in the halo-age--halo mass-gap relation can be attributed to the transitory nature of the halo mass-gap statistic. The halo mass-gap of a halo can evolve significantly and rapidly throughout the course of its evolution. Therefore, a single snapshot halo mass-gap (as is observed) can be a misleading representation of the true halo-age (see also \citealt{beckmann08}; \citealt{dariush10}; \citealt{cui11}). 

The transient nature of the satellite subhalo population can be further explored by using halo mass-gap as a rough indicator for the satellite subhalo merging timescale (see e.g. \citealt{boylan08}; \citealt{wetzel10}). This is the timescale for a satellite subhalo to go from virial infall to merging with the central subhalo:
\begin{equation}
T_{\rm merge}/T_{\rm hubble}=C_{\rm dyn} \frac{M_{\rm halo, inf}/M_{\rm sat,inf}}{\mathrm{ln}\left(1+M_{\rm halo, inf}/M_{\rm sat,inf}\right)}
\end{equation}

Here, $M_{\rm halo, inf}$ and $M_{\rm sat, inf}$ are the halo mass and satellite subhalo mass at the time of satellite infall, and $C_{\rm dyn} \sim 0.25$ is a free parameter empirically calibrated by \cite{wetzel10} to parameterize the satellite subhalo disruption rate. To convert the $z=0$ halo mass-gap to this timescale, we use the halo mass at infall rather than the present day halo mass. This has a negligible affect for high halo mass-gap systems but is important in the low halo mass-gap regime. The second $x$-axis on Fig. \ref{fig:dmass_age} shows that halo mass-gap systems with $\Delta M_h < 1$ have $T_{\rm merge}/T_{\rm hubble} < 1$ and are transient on average. Thus, a low halo mass-gap system can undergo a major merger within a Hubble time, and suddenly become a high halo mass-gap system (see path indicated by arrows). In addition, halo mass-gap is relatively insensitive to halo age when $\Delta M_h \gtrsim 1.5$ as $T_{\rm merge}/T_{\rm hubble} > 1$.

In Fig. \ref{fig:fractions} we show the fraction of ``Old'' ($T_{50} > 7.4$ Gyr), ``Intermediate'' ($4.3 < T_{50}/\mathrm{Gyr} < 7.4$) and ``Young'' ($ T_{50} < 4.3$ Gyr) halos as a function of halo mass-gap\footnote{Note here, and for the remainder of the paper, we use $T_{50}$ to describe ``halo age'', but our results are qualitatively unchanged if instead we adopt $T_{85}$.}. Here, we have defined intermediate-age halos as those which lie in the inter-quartile range of the overall distribution of halo ages, and young/old halos occupy the 25/75 percentiles respectively. The solid filled regions indicate halo mass-gap, while the line-filled regions are for stellar mass-gaps (see following section). As expected, the fraction of young/old halos decreases/increases with increasing halo (and stellar) mass-gap. However, as alluded to above, there remains a significant fraction of \textit{young} halos with large halo mass-gaps. Thus, the halo mass-gap statistic alone is insufficient to isolate truly old halos, and we caution against defining these extreme\footnote{Large stellar mass-gap systems are much rarer than smaller stellar mass-gap systems (see e.g. \citealt{yang08})} halo mass-gap systems as ``old and relaxed''.

\subsubsection{Physical or statistical relation?}
In the previous sections we have implicitly assumed that the relation between halo mass-gap and halo age is physical, namely that the halo mass-gap at $z=0$ contains information about the mass accretion history of the halo. However, it is worth noting that previous studies have shown that group \textit{richness} is also related to halo age at fixed halo mass (see e.g. \citealt{zentner05}). The more times that one draws from the stellar mass (or luminosity) function, the more likely it is to obtain a smaller halo mass-gap; therefore, there is a statistical relation between richness and halo mass-gap that is unrelated to the assembly history of the halo. Thus, at large/small mass-gaps we could simply be sampling the lowest/highest richness groups (see e.g. \citealt{paran12}; \citealt{hearin13a})

To address this issue we show in Fig. \ref{fig:dmass_age_nfixed} the relation between halo-age indicators ($T_{50}$ and $T_{85}$, cf. Fig. \ref{fig:dmass_age}) and halo mass-gap in fixed \textit{richness} bins. It is clear that the trend between halo mass-gap and halo age exists at fixed richness. Note that we also ensure that the relation holds at fixed halo mass \textit{and} fixed richness. The relation between richness and halo age at fixed halo mass-gap is also evident, but this is a much weaker trend that the halo age--halo mass-gap relation.

\subsection{Relation to observational quantities}
\label{sec:stellar}

\begin{figure}
  \begin{minipage}{\linewidth}
    \centering
    \includegraphics[width=8.5cm, height=4.7cm]{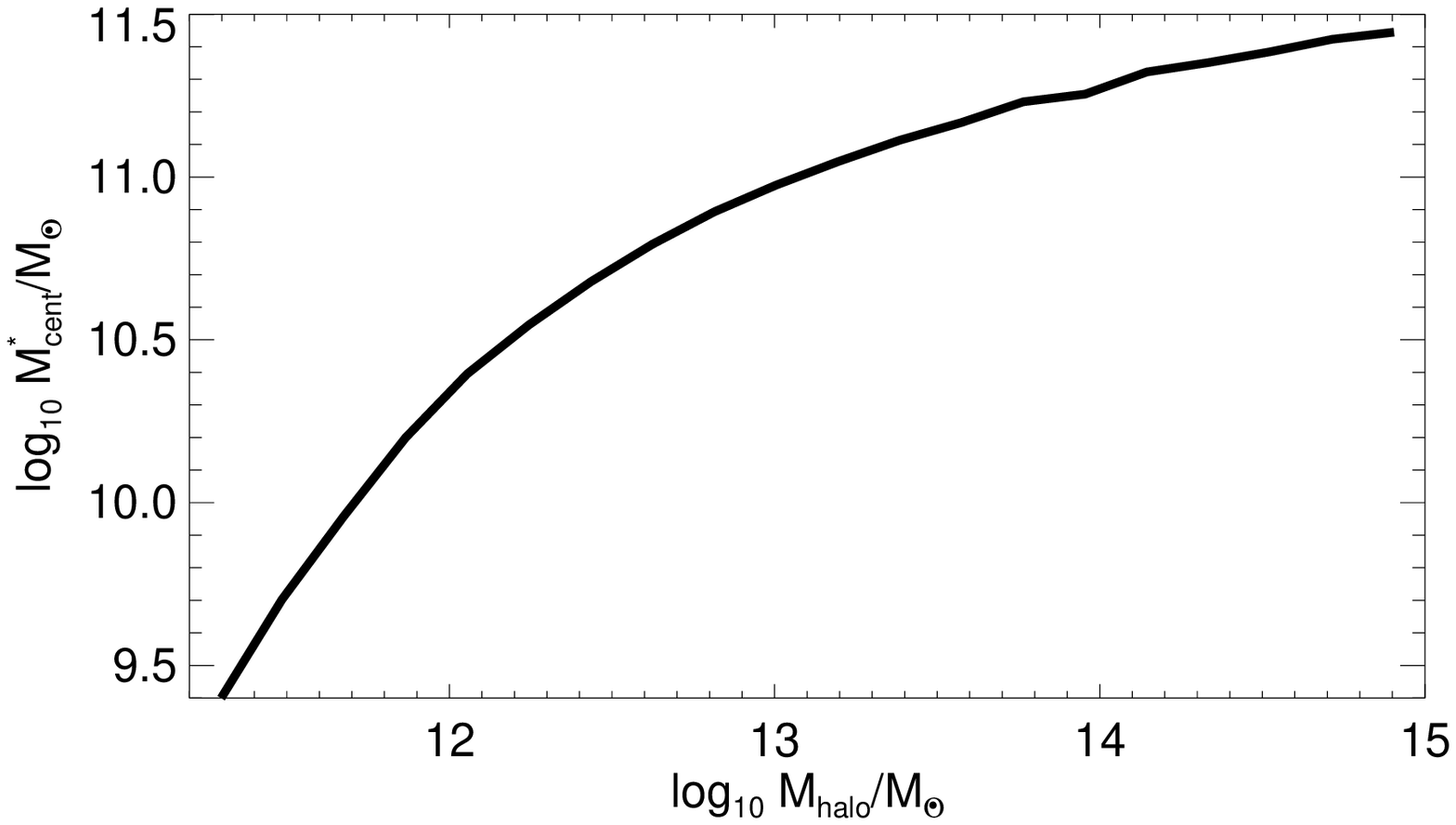}
   \end{minipage}
   \begin{minipage}{\linewidth}
     \vspace{-10pt}
     \centering
     \includegraphics[width=8.5cm, height=6.1cm]{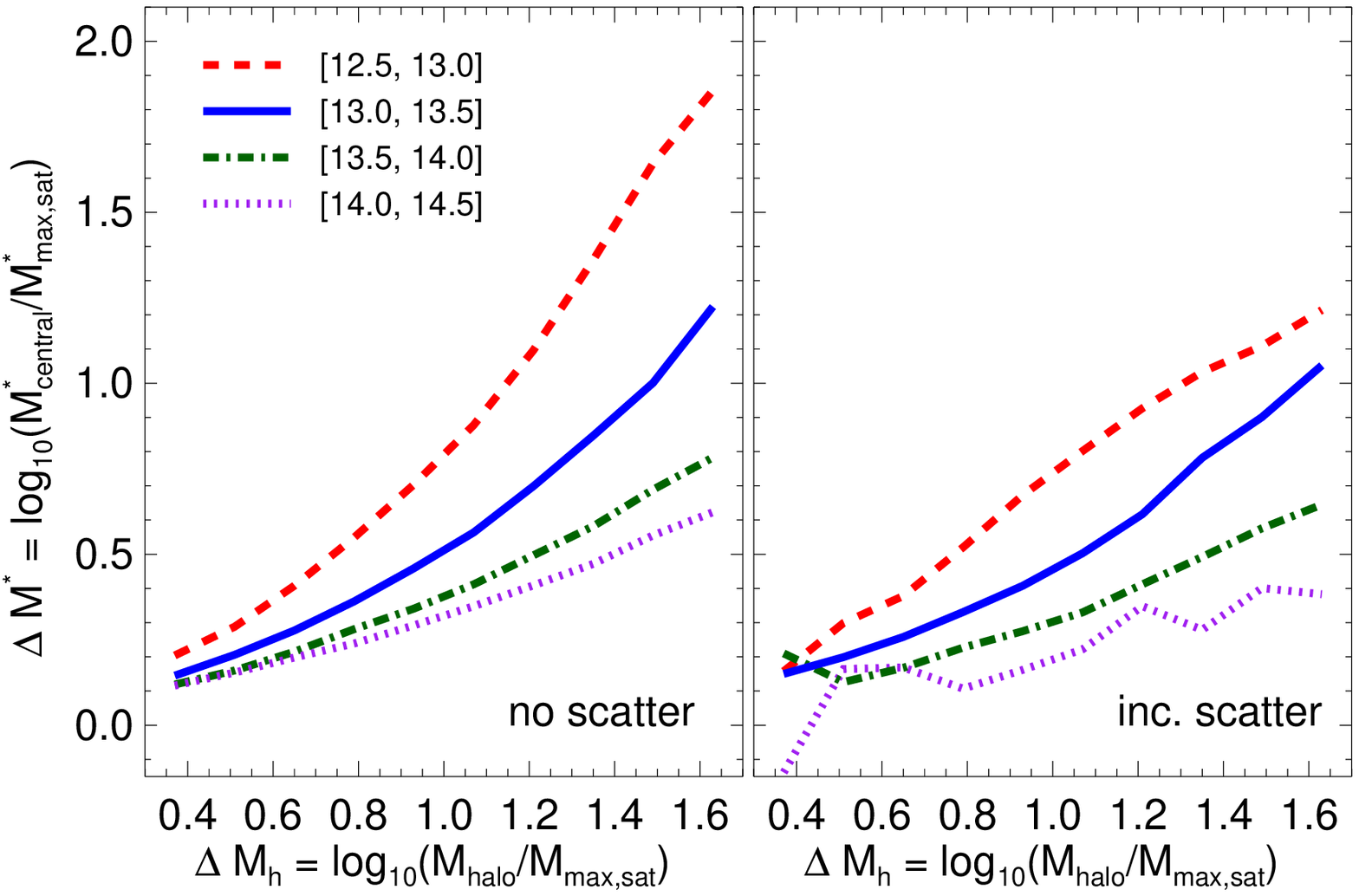}
   \end{minipage}
  \caption[]{\small \textit{Top inset:} The mean relation between central stellar mass and halo mass derived from abundance matching. At high masses the relation between stellar and halo mass flattens. \textit{Bottom panels:} The median relation between halo mass-gap and stellar mass-gap in four different halo mass bins. Stellar masses are assigned using halo abundance matching with the \cite{li09} stellar-mass function. In the right-hand panel a 0.15 dex scatter is included in the stellar mass-halo mass relation, and a lower limit of $M^*_{\rm sat} > 10^{9.7}M_\odot$.}
  \label{fig:dmhalo_dmstar}
\end{figure}

\begin{figure*}
    \centering
    \includegraphics[width=13.3cm, height=10cm]{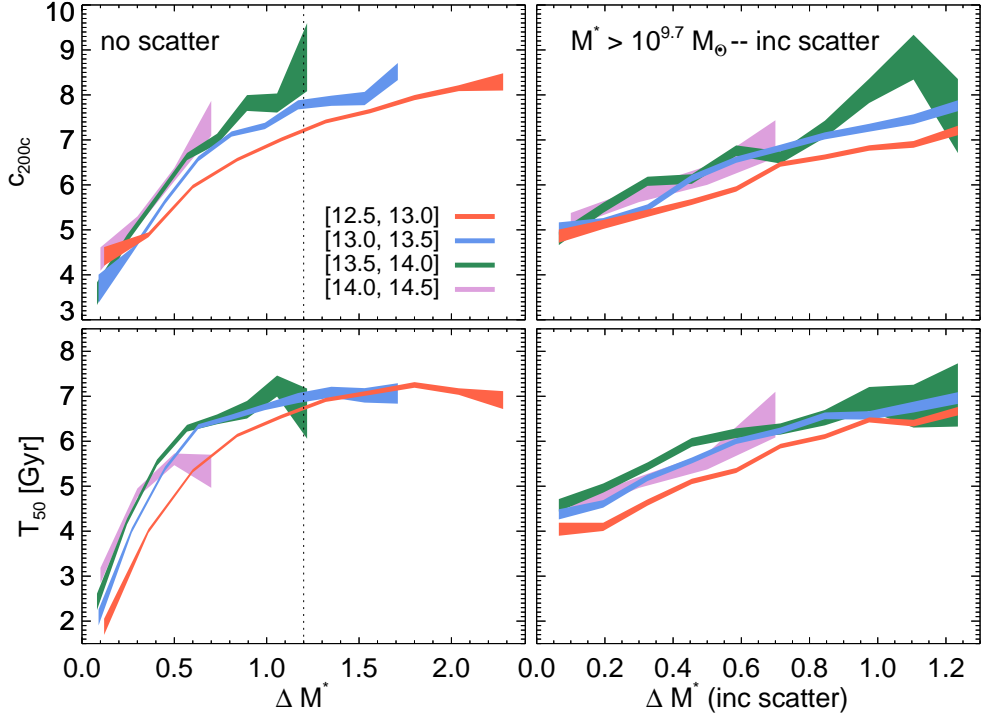}
  \caption[]{\small Halo concentration (top-panels) and halo age ($T_{50}$, bottom-panels) as a function of stellar mass-gap. The red, blue, green and purple shaded regions indicate the median values (and 1$\sigma$ uncertainty) for halos in different mass ranges. In the right-hand panels a 0.15 dex scatter is included in the stellar mass-halo mass relation, and we employ a lower mass cut of $M^* > 10^{9.7} M_\odot$. For stellar mass-gaps greater than $\Delta M^* \gtrsim 1.2$ there is very little correlation with halo concentration or age. The dotted line indicates this approximate limit whereby stellar mass-gap is a poor proxy for halo age/concentration.}
   \label{fig:conc_dmstar}
\end{figure*}

To relate halo mass-gap to an observational quantity we employ the widely used abundance-matching technique (e.g. \citealt{vale04}; \citealt{conroy06}; \citealt{behroozi10}; \citealt{guo10}; \citealt{moster10}). In brief, the halo mass function is mapped onto the observed stellar mass-function under the assumption that both these relations are monotonic (i.e. the most luminous galaxies are assigned to the most massive halos). The form of the relation between halo mass and stellar mass contains information about the efficiency of galaxy formation in different halo mass regimes. We use the stellar mass function for $z=0$ galaxies derived by \cite{li09} from the Sloan Digital Sky Survey (SDSS) Data Release 7 (DR7). This stellar mass function is based on the same stellar masses that we use in our group galaxy catalog (see Section \ref{sec:groups}). Stellar masses are assigned to subhalos using their ``peak'' mass ($M_{\rm peak}$) to account for the affects of tidal stripping (see e.g. \citealt{wetzel10}; \citealt{reddick13}).

In Fig. \ref{fig:dmhalo_dmstar} we show the resulting relation between halo mass-gap and stellar mass-gap. The dashed red ($10^{12.5} < M_{\rm halo}/M_\odot < 10^{13}$), solid blue ($10^{13} < M_{\rm halo}/M_\odot < 10^{13.5}$), dot-dashed green ($10^{13.5} < M_{\rm halo}/M_\odot < 10^{14}$) and dotted purple ($10^{14.0} < M_{\rm halo}/M_\odot < 10^{14.5}$) lines indicate the median trend for different halo mass bins. The relation between halo mass and stellar mass is weakest for the highest mass bin. This is because the form of the stellar mass-halo mass relation flattens at high masses, and halo mass becomes a progressively shallower approximation for stellar mass (see top inset panel of Fig. \ref{fig:dmhalo_dmstar}). In the right-hand panel we assign stellar mass via subhalo abundance matching assuming 0.15 dex log-normal scatter in $M^*$ at fixed $M_{\rm peak}$. This amount of scatter is motivated by observations, which suggest a scatter of 0.15-0.2 dex (e.g. \citealt{yang08}; \citealt{moster10}; \citealt{wetzel10}; \citealt{more11}). We also impose a stellar mass cut of $M^*_{\rm sat} > 10^{9.7}M_\odot$. Thus, the relation in the right-hand panel should more closely approximate the observational quantities of interest (see Section \ref{sec:obs}). 

Fig. \ref{fig:conc_dmstar} shows halo concentration (top-panels) and halo age ($T_{50}$, bottom panels) against stellar mass-gap. The right-hand panels include scatter and a mass threshold for the stellar masses. The red, blue, green and purple shaded regions show the median trends for the four different mass bins. Note that higher mass halos seem to have on average \textit{higher} concentrations than lower mass halos, this is contrary to the mass-concentration relation which predicts the opposite trend. There are two reasons for this apparent contradiction: (1) Our restriction to group halos where $N_{\rm sat} \ge 1$ biases against high concentration low mass halos (see Fig. \ref{fig:conc_dmass}), and (2) the correlation between stellar mass and halo mass is steeper for low mass halos, thus large halo mass-gap systems (which generally have higher halo concentrations) have very large stellar mass-gaps which are generally below the detection threshold of $M^* > 10^{9.7}M_\odot$. 

For stellar mass-gaps $\Delta M^* < 1.2$ there is a relation with halo age/concentration, but for very large stellar mass-gaps ($\Delta M^* > 1.2$) there is little or no correlation. This is partly due to the non-linear relationship between stellar mass-gap and halo mass-gap in this regime (see Fig. \ref{fig:dmhalo_dmstar}), but also because the relation between halo concentration/age and halo mass-gap flattens at very large halo mass-gaps. The inclusion of 0.15 dex scatter in the halo mass-stellar mass relation further weakens the correlation between halo age/concentration and stellar mass-gap. Indeed, the small range in median halo age ($T_{50} =4-7$ Gyr) and concentration ($c_{200c}=5-8$) over an order of magnitude in stellar mass-gap suggests that a large statistical sample of galaxy groups is required to make use of this relation.

\subsection{Mock group catalog}
\label{sec:mock}

\begin{figure*}
    \centering
    \includegraphics[width=16cm, height=8cm]{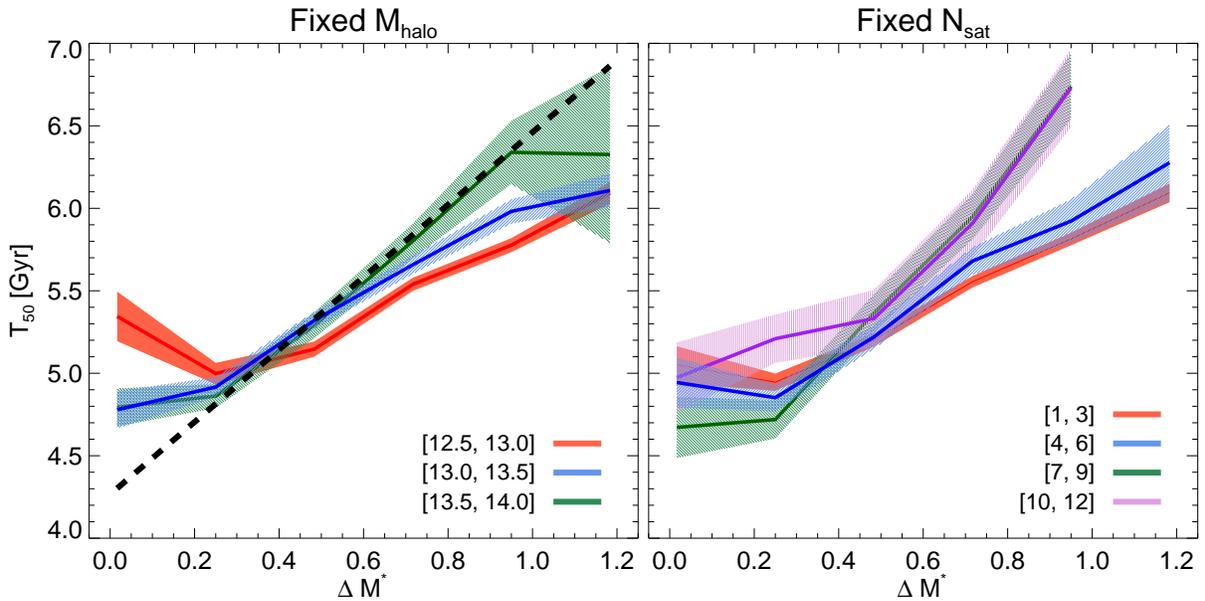}
  \caption[]{\small Halo age ($T_{50}$) as a function of stellar mass-gap in the mock group catalog. In the left-hand panel the relations for three halo mass bins are shown. The black dashed line shows the average relation seen in Fig. \ref{fig:conc_dmstar}, without applying the group catalog algorithm to the simulations. In the right-hand panel we show the relation in four richness bins (within the mass-range $10^{12.5} < M_{\rm halo}/M_{\odot} < 10^{14}$).}
   \label{fig:mock_test}
\end{figure*}

In the following section, we investigate the relation between galaxy properties and stellar mass-gap using a group catalog constructed from a local ($z \lesssim 0.04$) galaxy survey. Before proceeding, we investigate the affect of the group-finding algorithm (described below) on the halo age--stellar mass-gap relation using a ``simulation mock group catalog''. This mock catalog is described in more detail in \cite{wetzel12} and \cite{wetzel13}, and is constructed by applying the same group-finding algorithm to our simulations. This allows us to effectively ``observe'' our model results through this group catalog, which includes the effects of interloping galaxies caused by redshift-space distortions and any other observational systematics of the group-finding algorithm. The details of the group catalog are briefly described in Section \ref{sec:groups}.

In Fig. \ref{fig:mock_test} we show the relation between halo-age and stellar mass-gap in our mock group catalog. In the left-hand panel we consider three halo mass bins, where the mean relations and errors in the mean are indicated by the solid lines and shaded regions respectively. For comparison, the dashed black line shows the approximate relation between halo mass-gap and halo age shown in Fig. \ref{fig:conc_dmstar} which takes into account the scatter in the halo mass-stellar mass relation, but does not include the systematic effects of the group finding algorithm. It is encouraging to see that the group-finding algorithm does not wash out the halo age--stellar mass-gap relation. As perhaps expected, the trend is slightly weaker, and more so for lower mass-halos. In the right-hand panel we perform the same exercise but now consider four fixed group richness bins. The higher richness bins show a stronger relation than the low richness bins, presumably because the effect of interlopers is more important in the low-richness regime.

\section{Insights from observations}
\label{sec:obs}
\subsection{Group catalog}
\label{sec:groups}

Our galaxy sample is based on the NYU Value-Added Galaxy Catalog (\citealt{blanton05}) from SDSS DR7. The stellar masses are derived from the \textsc{kcorrect} code of \cite{blanton07a}, which assumes a \cite{chabrier03} initial mass function. A galaxy sample complete down to stellar masses $M^* > 10^{9.7}M_\odot$ is constructed, which consists of 21,423 local galaxies with redshifts $0.02 < z < 0.04$.

Groups of galaxies are identified using the procedure outlined in \cite{tinker11}, which is a modified implementation of the group finding algorithm in \cite{yang05, yang07}. Host halo masses are defined such that the mean matter density interior to the virial radius is 200 times the mean background matter density (i.e. $M_{\rm halo} = 200\rho_m \frac{4}{3} r^3_{\rm halo}$, cf. Section \ref{sec:sims}). Dark matter halo masses are assigned by matching the abundance of halos above a given dark matter mass to the abundance of groups above a given total stellar mass: $n(> M_{\rm halo})=n(> M^*)$. We use the host halo mass function from \cite{tinker08b}, which is based on a very similar cosmology to the simulations employed in the previous sections. Every group contains one ``central'' galaxy, which by definition is the most massive, and can contain any number of less massive ``satellite'' galaxies. We only consider groups with at least one satellite galaxy within $r_{\rm halo}$. The final sample contains $N=1240$ groups with halo masses in the range $10^{12.5} < M_{\rm halo}/M_\odot < 10^{14}$.

\cite{tinker11} investigate the robustness of the group catalog algorithm using mock galaxy distributions from simulations. These authors show that the completeness fractions for central/satellite galaxies in the relevant halo mass-range are approximately 95/80\%, and the purity fractions are 90/80\%. The group catalogs are described more extensively in \cite{tinker11}, \cite{wetzel12} and \cite{wetzel13}, and we refer the interested reader to these papers.

\subsection{(Non-) Relation between halo-age and galaxy-age}

\begin{figure*}
\centering
\includegraphics[width=18cm, height=18cm]{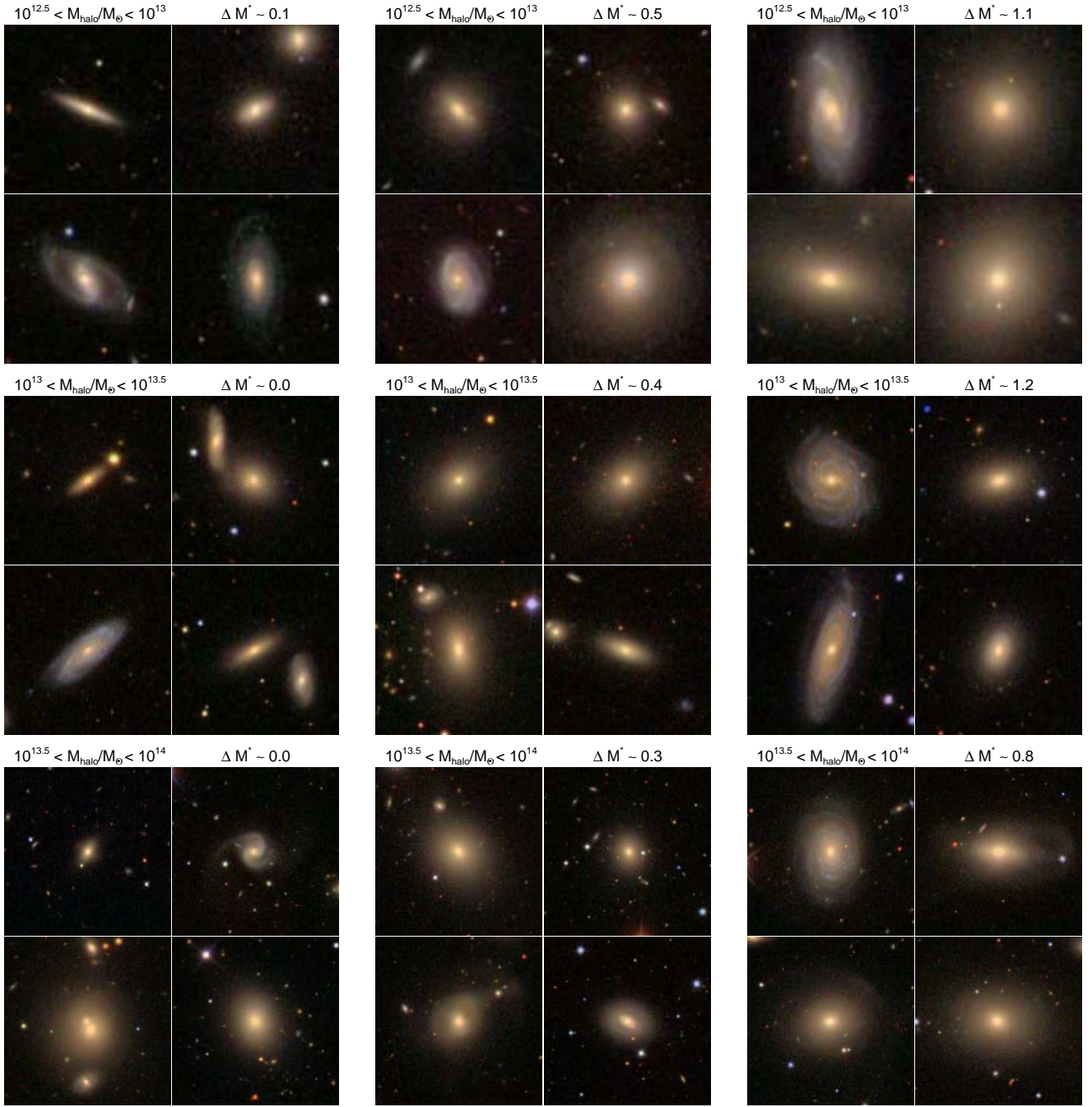}
\caption[]{\small Image montage of the central galaxies in the group catalog. Examples for low ($10^{12.5} < M_{\rm halo}/M_\odot < 10^{13}$), intermediate ($10^{13} < M_{\rm halo}/M_\odot < 10^{13.5}$), and high ($10^{13.5} < M_{\rm halo}/M_\odot < 10^{14}$) halo mass groups are shown in the top, middle and bottom panels respectively. The left, middle and right -hand panels show low, intermediate and high stellar mass-gaps respectively. The size of each panel is approximately $0.1 r_{\rm halo}$ for each mass bin, corresponding to $1.4, 2.1$ and $2.8$ arcmin for low, intermediate and high mass halos respectively.}
\label{fig:image_montage}
\end{figure*}

Old halos have accumulated most of their mass at early times, and have little recent mass growth, whereas old galaxies have stopped growing in stellar mass through star formation (often termed ``quenched'' galaxies). However, it is not clear whether or not old galaxies are preferentially found in old halos (see e.g. \citealt{blanton07b}; \citealt{tinker11}).

Here, we consider stellar mass-gap as a proxy for halo age. In the previous sections we noted the large scatter in the halo-age--mass-gap relation, which limits the use of stellar mass-gap as an age indicator on an individual object-by-object basis. However, for large group catalogs, such as the one under consideration here, we anticipate that \textit{average} trends may become apparent. In Fig. \ref{fig:image_montage} we show an image montage of central galaxies with different halo masses (increasing top to bottom) and stellar mass-gaps (increasing left to right). Here, the ``low'' and ``high'' stellar mass-gap systems make up the lowest or highest 10\% of stellar mass-gaps for each halo mass bin. We label the 10\% of halos centered on the median stellar mass-gap as the ``intermediate'' systems. For reference, the average stellar mass-gaps are approximately $\Delta M^* =0, 0.4, 1.0$ for the low, intermediate and high cases respectively. For each halo mass and stellar mass-gap bin, we show four example halos. These are drawn at random and a visual inspection ensures they are representative of the overall population. Note that the size of each panel is approximately $\sim 0.1 r_{\rm halo}$, so different mass halos have different angular sizes. For many of the low stellar mass-gap systems a massive satellite galaxy is visible, even on these small scales. Evidence of tidal interactions is also apparent, but there is no obvious trend between these features and stellar mass-gap.

The most striking aspect of this collection of images is the \textit{lack of any trend} in central galaxy properties with stellar mass-gap. One may naively expect that the relatively isolated, large stellar mass-gap systems would mainly consist of ``red and dead'' elliptical galaxies. However, as is evident in Fig. \ref{fig:image_montage}, a significant number of large stellar mass-gap systems have central galaxies which are spirals. This ``randomness'' in central galaxy properties is likely related to the large scatter in the halo-age--halo mass-gap relation we found in the preceding sections. However, we would expect at least an average trend to emerge if there was a strong relation between halo age and galaxy age.

The presence of relatively isolated, massive spirals in the group catalog is somewhat surprising, so we performed a number of sanity checks on these systems. First, a visual inspection of the fields surrounding the intermediate/high mass, large stellar mass-gap systems showed that the majority are indeed isolated. Second, we looked at the velocity distribution of the satellite members and found no significant difference between systems with a central spiral or early-type galaxy. The stacked velocity dispersion for intermediate (high) mass, large stellar mass-gap systems is $\sigma_{\rm early-types} \approx 200 (290)$  km s$^{-1}$ and $\sigma_{\rm spirals} \approx 160 (260)$ km s$^{-1}$ for central early-types and spirals respectively. The satellite velocity dispersions for large stellar mass-gap systems with a central spiral galaxy are slightly lower than those with a central early-type galaxy. If we assume $M_{\rm halo} \propto \sigma^3$ (see e.g. \citealt{nfw}), we find that the ratio of halo masses between systems with a central early-type or spiral galaxy varies by a factor of $\approx 1-2$. This difference in mass is not significant enough to suggest that the halo masses assigned to these extreme systems by the group catalog algorithm are notably spurious.  

\begin{figure*}
  \begin{minipage}{\linewidth}
    \centering
    \includegraphics[width=14cm, height=10cm]{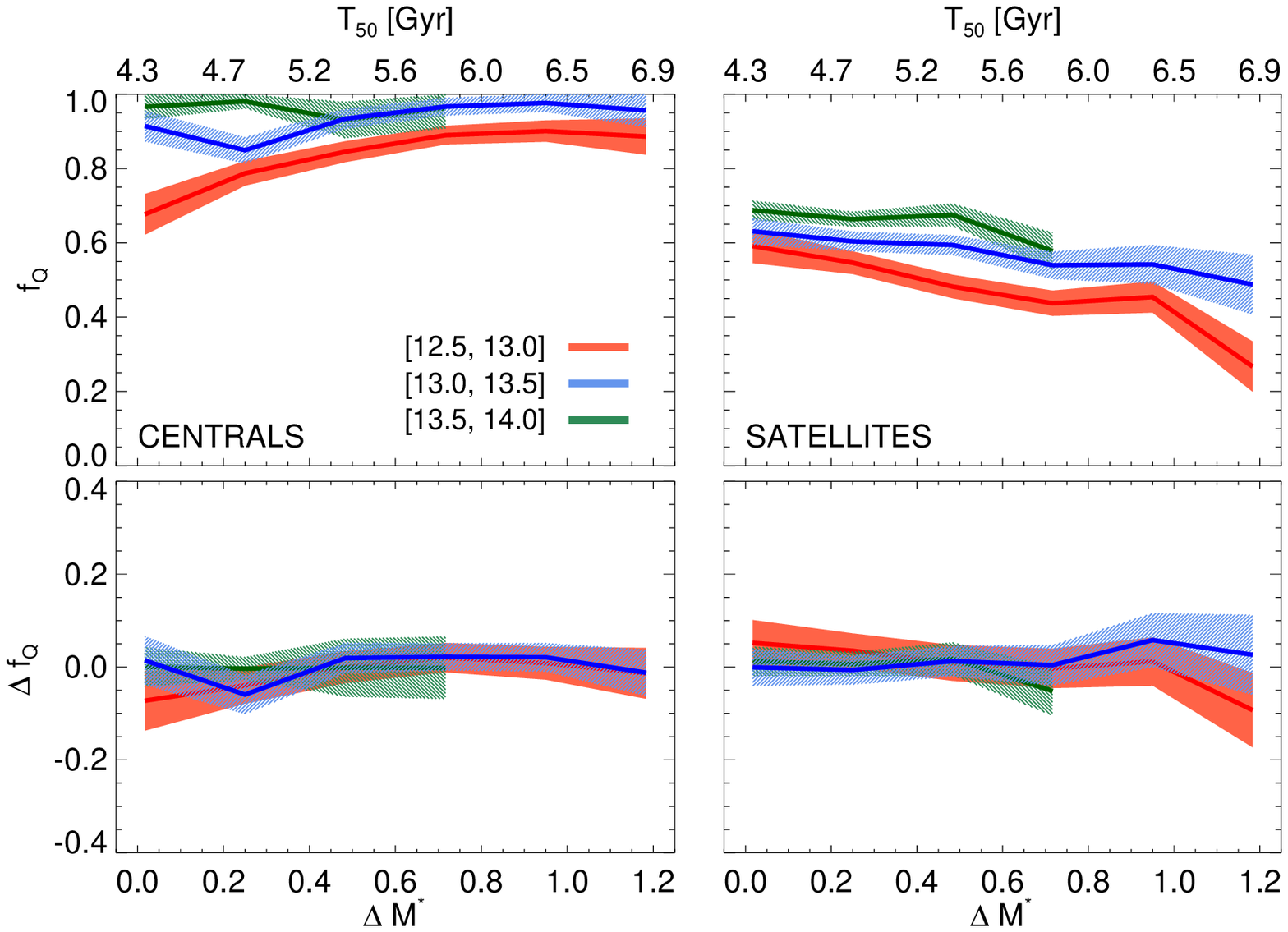}
    \caption[]{\small The relation between the quenched fraction of galaxies ($f_Q$) and stellar mass-gap. The solid lines show the mean values, and the shaded regions indicate the associated error in the mean. The top-axis shows the approximate halo formation time ($T_{50}$). This is calculated from the roughly linear relation between $T_{50}$ and stellar mass-gap shown in the bottom-right hand panel of Fig. \ref{fig:conc_dmstar}. The relations for central/satellite galaxies are shown in the left/right hand panels respectively. The bottom panels show the relations when the stellar mass-dependence is removed. Thus, when the stellar mass-dependence is taken into account, there is little correlation between galaxy age and stellar mass-gap.}
  \label{fig:fq_props}
  \end{minipage}
  \begin{minipage}{\linewidth}
    \vspace{10pt}
    \centering
    \includegraphics[width=14cm, height=10cm]{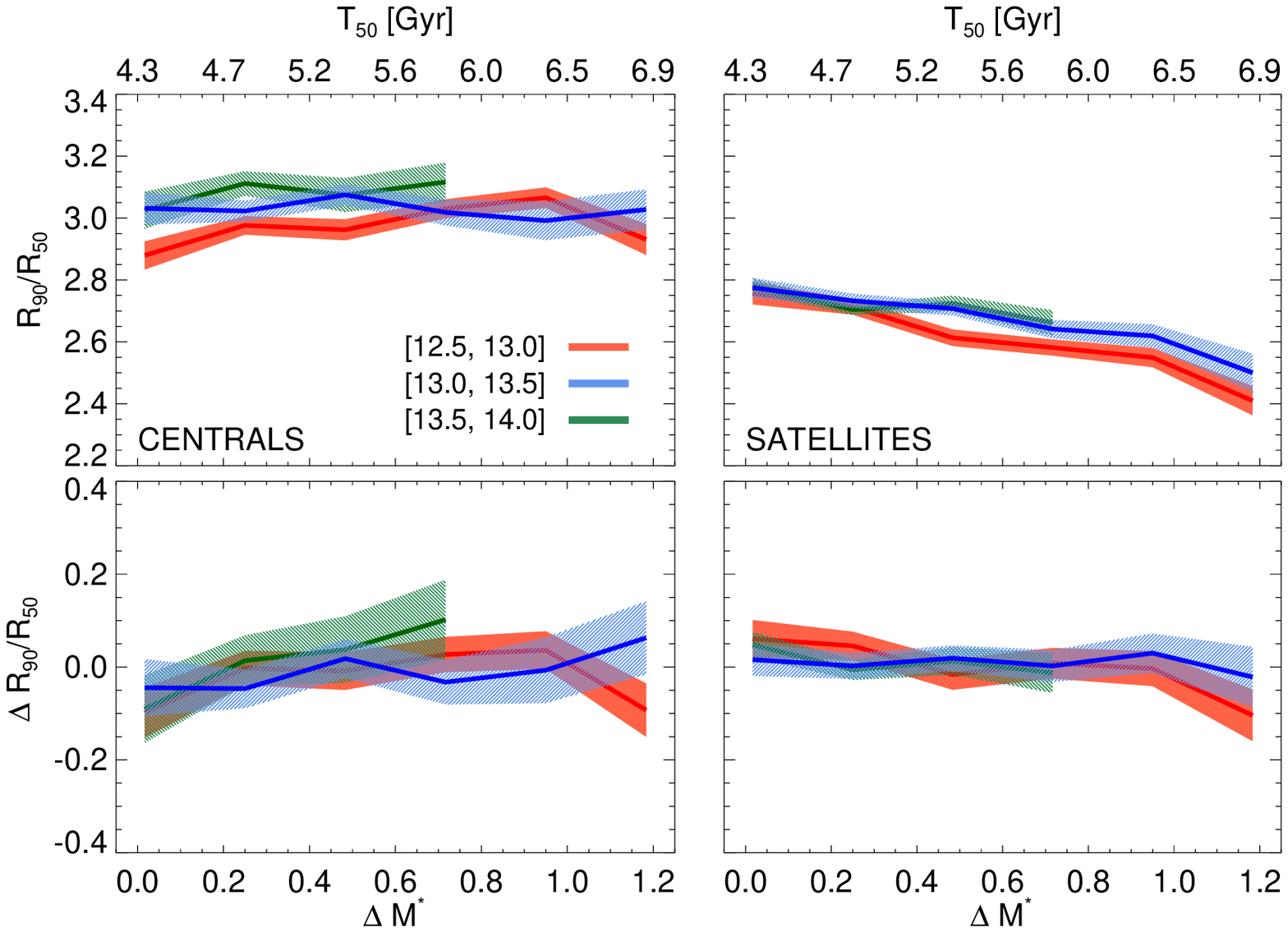}
  \end{minipage}
  \caption[]{\small Same as above figure, but here we show a measure of galaxy concentration -- the ratio between 90\% and 50\% Petrosian radii --  as a function of stellar mass-gap. At fixed halo/stellar mass there is no correlation between stellar mass-gap and galaxy concentration for both satellites and centrals.}
  \label{fig:cc_props}
\end{figure*}

In Fig. \ref{fig:fq_props} we investigate the relation between galaxies and stellar mass-gap more quantitatively. We use the specific star formation rate, SSFR=SFR/$M^*$, to describe the galaxy star formation. This metric is based on the current release of the spectral reductions of \cite{brinchmann04}. We follow the definition of \cite{wetzel13} and define galaxies with SSFR $< 10^{-11}$ yr$^{-1}$ as ``quenched''. Note that we also consider the 4000-\AA\ break, $D_n 4000$, as a star formation metric, where $D_n 4000$ is a diagnostic of the light-weighted age of the stellar population. Here, we follow \cite{tinker11} and label galaxies with $D_n 4000 > 1.6$ as quenched. In the following we only use SSFR to define quenched galaxies, but we find very similar trends if the $D_n 4000$ diagnostic is used instead.

In the top panels of Fig. \ref{fig:fq_props} we show the quenched fraction ($f_Q$) of galaxies against stellar mass-gap for centrals (left panels) and satellites (right panels). Low ($10^{12.5} < M_{\rm halo}/M_\odot < 10^{13}$), intermediate ($10^{13} < M_{\rm halo}/M_\odot < 10^{13.5}$) and high ($10^{13.5} < M_{\rm halo}/M_\odot < 10^{14}$) mass halos are shown with the red, blue and green filled regions respectively. The mean relation is indicated with the solid line, and the filled regions show the uncertainty in the mean.

At first glance it appears that the fraction of quenched galaxies is related to stellar mass-gap, where the $f_Q$ of centrals/satellites increases/decreases with increasing stellar mass-gap. However, by virtue of the group catalog algorithm, where halo mass is assigned based on \textit{total} stellar mass,  higher stellar mass-gap systems have higher central stellar masses and lower satellite stellar masses. Thus, the trends in the top panels of Fig. \ref{fig:fq_props} could be entirely driven by the correlation between $f_Q$ and galaxy stellar mass at fixed halo mass. To account for this, we compute in each stellar mass-gap bin the average stellar mass of centrals or satellites, and find the average $f_Q$ for galaxies at these stellar masses. In the bottom panels of Fig. \ref{fig:fq_props} we remove any stellar mass dependence by showing the difference in  $f_Q$ ($\Delta f_Q$) between galaxies at fixed stellar mass-gap and galaxies at fixed stellar mass. We find that for each mass bin both centrals and satellites have $\Delta f_Q$ close to zero, which indicates the trends in the top panels are indeed driven by stellar mass.

In Fig. \ref{fig:cc_props} we consider the structural properties of the central and satellite galaxies. Here, we use the ratio between the 90\% and 50\% Petrosian radii to denote galaxy concentration. In a similar fashion to the star formation properties of the galaxies, we find no apparent trend between galaxy concentration and stellar mass-gap at fixed mass. Given the relatively strong trend between halo concentration and stellar mass-gap (see Fig. \ref{fig:conc_dmstar}), this suggests that there is little relation between the ``compactness'' of galaxies and their halos.

In Fig. \ref{fig:conc_dmstar} we showed that there is an approximately linear relation between halo formation time and stellar mass-gap. We estimate this average linear relation for halo masses in the range $10^{12.5} < M_{\rm halo}/M_\odot < 10^{14}$. The resulting relation ($T_{50} \approx 4.3 +2.2\Delta M^*$) is used to give the approximate halo formation time appropriate for each stellar mass-gap bin, and this is shown in the top $x$-axis labels of Fig. \ref{fig:fq_props} and Fig. \ref{fig:cc_props}. Over the range of stellar mass-gaps from $\Delta M^*=0$ to $\Delta M^* =1.2$ the formation time changes by $\sim 2.5$ Gyr ($T_{50} \approx 4.5-7$ Gyr). This suggests that a variation in $T_{50}$ by $\approx 40\%$ results in no appreciable difference in the star formation or structural properties of galaxies at fixed mass.

This more quantitative assessment of the relation between galaxy properties and stellar mass-gap is in agreement with the visual interpretation that we found in Fig. \ref{fig:image_montage}. We find little or no relation between galaxies and stellar mass-gap. We also consider excluding low-richness systems which are probably more affected by interlopers (with $N \lesssim 3$ group members, see Fig. \ref{fig:mock_test}), but we still see no evidence for a relation between galaxy properties and stellar mass-gap when we are restricted to higher richness systems. Given the degree of scatter in the intrinsic relation between stellar mass-gap and halo age, we cannot discount a weak residual trend exists. However, these findings are inconsistent with a \textit{strong} relation between galaxy age and halo age. Larger and/or deeper samples of galaxies, such as the NASA-Sloan Atlas (NSA), zCOSMOS and the Galaxy And Mass Assembly (GAMA) survey, will help to add more statistical weight to these findings.

\section{Discussion}

\subsection{Implications for Milky Way mass halos and fossil groups}

The use of stellar mass-gap as a halo age indicator is strongly dependent on halo mass. Fig. \ref{fig:dmhalo_dmstar} shows that the relation between stellar mass-gap and halo mass-gap gets progressively shallower with increasing halo mass. This means that for large mass halos ($M_{\rm halo} \gtrsim 10^{14}M_\odot$), a large change in halo mass-gap, and therefore approximately halo-age, translates to a small change in stellar mass-gap. This small dynamic range may limit our ability to disentangle old and young halos in high mass systems. On the other hand, the limiting factor for lower mass halos is likely the depth of the galaxy survey. However, we can address this limitation by probing fainter satellite galaxies. In this work, we consider group/cluster mass halos with $M_{\rm halo} > 10^{12.5} M_\odot$, however, stellar mass-gap could potentially be a powerful tool for Milky Way mass halos ($M_{\rm halo} \sim 10^{12} M_\odot$). On lower halo mass scales, deeper samples of galaxies will be vital in order to probe a significant portion of the satellite luminosity function. Especially since large stellar mass-gap systems are more common at lower mass scales (cf. the Milky Way and M31 both have $\Delta M^* \sim 1$). Fortunately, there are several upcoming galaxy surveys (mentioned above) which will provide large samples of galaxies down to much lower stellar masses. 

On the group mass scales considered in this work, our results may have important ramifications for studies of ``fossil groups''. This class of groups are, by definition, X-ray luminous galaxy groups dominated by one massive elliptical galaxy central with a dearth of massive companion satellites. Our finding that $\sim 20$ \% of large halo mass-gap systems may have recently experienced a major merger is contrary to the expectation that fossil groups are ``old and relaxed'' (see also \citealt{beckmann08} and \citealt{dariush10}). This latter point is important as fossil groups are often claimed to be useful test-beds for $\Lambda$CDM cosmology owing to their apparently virialized state. Furthermore, the very definition of fossil groups ignores the non-negligible fraction of isolated, massive groups dominated by a giant spiral galaxy. These systems may be particularly interesting, as the presence of a central spiral galaxy argues against a recent merger with a massive satellite. Are these massive, isolated disks examples of truly old galaxy halos? Further investigation into these systems will help to address this question.

\subsection{Galaxy assembly bias}
The properties of dark matter halos, such as their formation times and merger histories, are strongly correlated with halo mass and environment (e.g., \citealt{gao05}; \citealt{harker06}; \citealt{wechsler06}; \citealt{wetzel07}; \citealt{dalal08}). However, the effect on the galaxy population is less clear. Models of the Halo Occupation Distribution (HOD) assume that galaxy occupation of dark matter halos depends on halo mass only. This standard implementation of the HOD has been called into question by a number of ``galaxy assembly bias'' models, in which galaxy color is correlated with the large-scale environment at fixed halo mass. The observational evidence for or against galaxy assembly bias is conflicting. \cite{yang06} split 2dF galaxy groups by mass and SFR and find that the clustering of groups decreases with increasing SFR at fixed mass (see also \citealt{wang08}). However, \cite{berlind06} find the opposite trend in SDSS galaxy groups, whereby blue central galaxies are more clustered than red central galaxies at the same mass. Furthermore, several studies have found that the color dependent clustering of SDSS galaxies are well fit by HOD models without dependence on large scale environment (e.g. \citealt{abbas06}; \citealt{blanton07b}; \citealp{tinker08a, tinker11}; \citealt{skibba09})

In this work, we find no evidence to suggest that the star formation properties of galaxies are related to halo age, in agreement with studies claiming that there is no significant galaxy assembly bias. These results are encouraging for the convenient HOD and abundance matching methods, in which the affects of environment on the galaxy occupation of dark matter halos are largely ignored. Finally, we note that stellar mass-gap dependent galaxy clustering may be an additional powerful tool in which to investigate galaxy assembly bias.

\section{Conclusions}
We consider halo mass-gap (and stellar mass-gap) as a proxy for halo age and concentration in group/cluster scale halos. From $z=0$ simulations, we show that halo age and concentration are related to the difference in logarithmic mass between the parent halo and its most massive satellite subhalo, but there is a significant amount of scatter which limits the use of halo mass-gap as an age indicator on an individual object-by-object basis. Our main conclusions are summarized as follows:

\begin{itemize}
\item The relation between halo age/concentration and halo mass-gap is related to simple dynamical arguments: the relatively short dynamical timescales of massive satellite subhalos means that early forming halos are likely devoid of a massive companion. We show that, over a similar dynamic range in halo mass or halo mass-gap, the correlation between halo mass-gap and concentration is stronger than the mass-concentration relation, and will likely prove a useful test of the $\Lambda$CDM paradigm. We find that the association between halo mass-gap and concentration can severely bias the mass-concentration relation of group catalogs; in stellar mass limited samples, low halo mass groups are biased against high concentration (and hence high halo mass-gap) systems, whereas higher mass halos are relatively unaffected. This can lead to an seemingly flat mass-concentration relation if this bias is not accounted for.

\item On average, larger halo mass-gap systems (up to a limit of $\Delta M_h \sim 1.5$) have older and more concentrated halos. However, we find that there is a significant amount of scatter in the halo-age/concentration--halo mass-gap relation that increases at larger halo mass-gaps. This scatter reflects the transitory nature of the halo mass-gap statistic. We find that a single redshift snapshot of the halo mass-gap is insufficient to describe the true age of the halo. For example, a significant number ($\sim 20\%$) of large halo mass-gap systems are relatively young ($T_{50} \lesssim 4$ Gyr) systems whereby there has been a recent merger between a massive satellite subhalo and the central subhalo.

\item  We relate halo mass-gap to stellar mass-gap using abundance matching. Owing to the flat relation between stellar mass and halo mass at the high mass end, stellar mass-gap is most useful as a halo age indicator for lower mass group halos ($M_{\rm halo} \lesssim 10^{12.5}M_\odot$). The relation between halo age/concentration and stellar mass-gap is weakened by the introduction of an observationally motivated amount of scatter and a stellar mass limit. However, we note that even after accounting for the observational sources of scatter, the correlation between stellar mass-gap and halo concentration is stronger than the mass-concentration relation.

\item Using a group catalog derived from the SDSS DR7 galaxy sample we investigate the relation between galaxy properties and stellar mass-gap. We find no significant relation between central or satellite galaxies and stellar mass-gap. This provides further evidence for the lack of correlation between galaxy age (or any other property) and halo age. However, given the large scatter we cannot reject the possibility that a weak residual trend exists. An inspection of the SDSS images shows a strikingly large range of central properties for very large stellar mass-gap systems. In particular, spiral galaxies with large stellar mass-gaps are not uncommon. This finding calls into question the commonplace definition of large stellar mass-gap systems as ``fossil groups''.

\end{itemize}

\section*{Acknowledgments}
AJD is currently supported by NASA through Hubble Fellowship grant HST-HF-51302.01, awarded by the Space Telescope Science Institute, which is operated by the Association of Universities for Research in Astronomy, Inc., for NASA, under contract NAS5-26555. CC acknowledges support from the Alfred P. Sloan Foundation. We thank Andrew Hearin, Tomer Tal and Andrew Zentner for useful comments and advice. We also thank the anonymous referee, whose comments greatly improved the quality of this paper.

\end{document}